\RequirePackage{latexrelease}
\documentclass[twocolumn,english,superscriptaddress,citeautoscript,preprintnumbers,amsmath,amssymb,prb,floatfix,footinbib]{revtex4-2}
\usepackage{hyperref}
\hypersetup{colorlinks=true,linkcolor=red,citecolor=blue}
\usepackage[scaled=2]{helvet}
\usepackage[T1]{fontenc}
\usepackage[latin9]{inputenc}
\usepackage{array}
\usepackage{float}
\usepackage{multirow}
\usepackage{amsmath}
\usepackage{amssymb}
\usepackage{graphicx}
\usepackage{csquotes}
\usepackage{xcolor,soul}

\makeatletter

\@ifundefined{textcolor}{}
{%
	\definecolor{BLACK}{gray}{0}
	\definecolor{WHITE}{gray}{1}
	\definecolor{RED}{rgb}{1,0,0}
	\definecolor{GREEN}{rgb}{0,1,0}
	\definecolor{BLUE}{rgb}{0,0,1}
	\definecolor{CYAN}{cmyk}{1,0,0,0}
	\definecolor{MAGENTA}{cmyk}{0,1,0,0}
	\definecolor{YELLOW}{cmyk}{0,0,1,0}
}

\usepackage[scaled=2]{helvet}\usepackage{amstext}

\makeatletter

\usepackage{dcolumn}
\usepackage{bm}
\usepackage{times}

\makeatother

\usepackage{babel}

\makeatother
\usepackage{babel}
\begin{document}
	\title{Weyl points and anomalous transport effects tuned by the Fe doping in Mn$_3$Ge Weyl semimetal}
	\author{V. Rai}
	\affiliation{Forschungszentrum J\"ulich GmbH, J\"ulich Centre for Neutron Science (JCNS-2)
		and Peter Gr\"unberg Institut (PGI-4), JARA-FIT, 52425 J\"ulich, Germany}
	\affiliation{RWTH Aachen, Lehrstuhl f\"ur Experimentalphysik IVc, J\"ulich-Aachen Research
		Alliance (JARA-FIT), 52074 Aachen, Germany}
	\affiliation{Department of Physics and Materials Science, University of Luxembourg, 162A avenue de la Faiencerie, L-1511 Luxembourg, Grand Duchy of Luxembourg}
	\author{S. Jana}
	\affiliation{Forschungszentrum J\"ulich GmbH, J\"ulich Centre for Neutron Science (JCNS-2)
		and Peter Gr\"unberg Institut (PGI-4), JARA-FIT, 52425 J\"ulich, Germany}
	\affiliation{RWTH Aachen, Lehrstuhl f\"ur Experimentalphysik IVc, J\"ulich-Aachen Research
		Alliance (JARA-FIT), 52074 Aachen, Germany}
	\author{J. Per\ss{}on}
	\affiliation{Forschungszentrum J\"ulich GmbH, J\"ulich Centre for Neutron Science (JCNS-2)
		and Peter Gr\"unberg Institut (PGI-4), JARA-FIT, 52425 J\"ulich, Germany}

	\author{S. Nandi}
	\email{s.nandi@fz-juelich.de}
	
	\affiliation{Forschungszentrum J\"ulich GmbH, J\"ulich Centre for Neutron Science (JCNS-2)
		and Peter Gr\"unberg Institut (PGI-4), JARA-FIT, 52425 J\"ulich, Germany}
	\affiliation{RWTH Aachen, Lehrstuhl f\"ur Experimentalphysik IVc, J\"ulich-Aachen Research
		Alliance (JARA-FIT), 52074 Aachen, Germany}

	\begin{abstract}
The discovery of a significantly large anomalous Hall effect in the chiral antiferromagnetic system - Mn$_3$Ge - indicates that the Weyl points are widely separated in phase space and positioned near the Fermi surface. In order to examine the effects of Fe substitution in Mn$_3$Ge on the presence and location of the Weyl points, we synthesized (Mn$_{1-\alpha}$Fe$_{\alpha})$$_3$Ge ($\alpha=0-0.30$) compounds. The  anomalous Hall effect was observed in compounds up to $\alpha=0.22$, but only within the temperature range where the magnetic structure remains the same as the Mn$_3$Ge. Additionally, positive longitudinal magnetoconductance and planar Hall effect were detected within the same temperature and doping range. These findings strongly suggest the existence of Weyl points in (Mn$_{1-\alpha}$Fe$_{\alpha})$$_3$Ge ($\alpha=0-0.22$) compounds. Further, we observed that with an increase in Fe doping fraction, there is a significant reduction in the magnitude of anomalous Hall conductivity, planar Hall effect, and positive longitudinal magnetoconductance, indicating that the Weyl points move further away from the Fermi surface. Consequently, it can be concluded that suitable dopants in the parent Weyl semimetals have the potential to tune the properties of Weyl points and the resulting anomalous electrical transport effects.

	\end{abstract}
	\maketitle

	\section{Introduction:}
	
The field of condensed matter physics has entered a new era with the recent discovery of topological phases of matter \cite{ando2013topological, hasan2010topological}. A significant breakthrough in this direction is the identification of chiral antiferromagnets that host Weyl Fermions \cite{nayak2016large, kuroda2017evidence}. The presence of Weyl nodes in chiral antiferromagnets leads to anomalous transport effects, which are rarely observed in conventional antiferromagnets, as mentioned in Refs. \mbox{\cite{nandy2017chiral, chen2021anomalous, yang2017topological}}. The observation of a substantial anomalous Hall effect (AHE) in chiral antiferromagnets is interesting, paving the way for research in the realms of topological materials, spintronics, and high-efficiency energy harvesting devices \cite{kuroda2017evidence, nayak2016large, higo2018large,ikhlas2017large, wuttke2019berry, tsai2020electrical, jungwirth2016antiferromagnetic, jeon2021long}. The separation of Weyl points in phase space and their positioning relative to the Fermi energy play a crucial role in determining the magnitude of AHE and the electrical transport effects induced by the chiral anomaly \cite{vsmejkal2018topological,wu2023temperature, wuttke2019berry, nandy2017chiral, yang2017topological, burkov2014chiral, burkov2016z, burkov2017giant}.

The chiral antiferromagnetic hexagonal phase of Mn$_3$Ge hosts multiple pairs of Weyl Fermions situated in proximity to the Fermi surface \cite{soh2020ground, yang2017topological}. As a result, Mn$_3$Ge  exhibits large anomalous Nernst effect, magneto-optical Kerr effect, AHE, and positive longitudinal magnetoconductivity (LMC), all justifying the presence of Weyl points near the Fermi surface \cite{rai_mn3ge, wuttke2019berry, wu2020magneto, nayak2016large, kiyohara2016giant, chen2021anomalous, jeon2021long}. Temperature-dependent Nernst effect measurements on Mn$_3$Ge have demonstrated that the position of Weyl points can be manipulated by changing the temperature \cite{wuttke2019berry}. Recent efforts have focused on controlling the Weyl nodes in various compounds through doping, external pressure, and other means \cite{liu2020evolution, zhou2020enhanced, sukhanov2018gradual, dos2020pressure, thakur2020intrinsic, ghimire2019creating, hu2021evolution,ikhlas2022piezomagnetic, low2022tuning}. However, the precise understanding of the role played by external factors in the dynamics of Weyl Fermions is still incomplete.

Our research focuses on investigating the characteristics of Weyl Fermions in chiral antiferromagnets by studying the evolution of Weyl points and the resulting electronic transport effects in hexagonal-(Mn$_{1-\alpha}$Fe$_{\alpha}$)$_3$Ge. While the evolution of magnetism and trivial transport effects with Fe doping in hexagonal-Mn$_3$Ge has been studied in the past \cite{lecocq1963, kanematsu1967, hori1992antiferromagnetic, hori1995magnetic, niida1995magnetic}, their recognition as potential Weyl semimetals has been reported recently \cite{rai_fe_neutron, achintyalow_Fe}. Previous studies \cite{hori1992antiferromagnetic, hori1995magnetic, niida1995magnetic} have shown that for low Fe doping levels ($\lesssim15\%$), the compounds exhibit magnetization similar to Mn$_3$Ge, even at the lowest measured temperature. However, it was observed that the N\'eel temperature ($T_{\text{N1}}$) decreases as Fe doping in Mn$_3$Ge increases. For higher Fe doping levels (approximately $0.15\lesssim\alpha\lesssim0.25$), a second magnetic transition appears at $T_{\text{N2}}$, which remains lower than $T_{\text{N1}}$. The magnetization of the compounds in these two magnetic regimes is distinct, indicating different magnetic structures in each temperature regime. Finally, in the case of even higher Fe doping levels ($\alpha\gtrsim0.30$), both antiferromagnetic transitions are suppressed by the emergence of ferromagnetism in the sample.

In this report, we initially conducted a comprehensive analysis of the magnetization properties of hexagonal-(Mn$_{1-\alpha}$Fe${_\alpha}$)$_3$Ge compounds with varying Fe doping fractions ($\alpha$) ranging from 0 to 0.30. Notably, we observed that the magnetization behavior resembling Mn$_3$Ge persists over a considerable temperature range, specifically up to $\alpha$ = 0.22. For compounds with $\alpha$ = 0.14 - 0.22, a change in magnetization behavior at low temperatures allowed us to investigate the influence of such changes on the electrical transport effects of the samples. Consequently, we performed a detailed analysis of the AHE and various magnetoconductivity (MC) measurements on single-crystal (Mn$_{1-\alpha}$Fe$_{\alpha}$)$_3$Ge compounds with $\alpha$ = 0 - 0.22. For low Fe doping levels ($\alpha$ = 0.04, 0.10), the AHE and positive LMC were observed, exhibiting similarities (albeit with lower magnitudes) to the parent sample, below the Ne\'el temperature ($T_{\text{N1}}$). As the Fe doping increased ($\alpha$ = 0.18, 0.22), the samples underwent a second magnetic transition at $T_{\text{N2}}$ (which is lower than $T_{\text{N1}}$). In these cases, the AHE and positive LMC were observed between $T_{\text{N2}}$ and $T_{\text{N1}}$. Moreover, it was found that these compounds exhibited a magnetic structure similar to Mn$_3$Ge between $T_{\text{N2}}$ and $T_{\text{N1}}$ \cite{rai_fe_neutron}. Interestingly, below $T_{\text{N2}}$, the magnetic structure changed to a collinear antiferromagnetic (AFM) configuration, and the AHE vanished. This observation suggests that the Weyl points exist in (Mn$_{1-\alpha}$Fe$_{\alpha}$)$_3$Ge compounds solely within the temperature range of $T_{\text{N2}}$ to $T_{\text{N1}}$.
	
The chiral anomaly effect in Weyl semimetals is characterized by several phenomena such as positive LMC, planar Hall effect (PHE), and angular magnetoconductivity ($\theta$MC) \cite{burkov2014chiral, burkov2017giant}. However, these effects can arise due to other effects as well, for example, due to the magnetization, or unequal spin density of states (sDOS) near the Fermi surface. Our analysis suggests that the observed positive LMC (at $B=1$ T), $\theta$MC, and PHE in the Fe-doped Mn$_3$Ge compounds are likely a result of the chiral anomaly effect in Fe-doped Mn$_3$Ge. However, further investigations are necessary to support our claim.

We have observed that the AHE and signatures of the chiral anomaly effect (positive LMC, $\theta$MC, and PHE) are present in the compounds with Fe doping levels ranging from $\alpha = 0$ to $\alpha = 0.22$, but only within the temperature range where the magnetic structure remains similar to that of Mn$_3$Ge. This suggests that the Weyl points exist in the Fe-doped samples as long as they retain a magnetic structure similar to Mn$_3$Ge. Moreover, the magnitude of AHE and LMC decreases with an increase in the Fe doping level ($\alpha$), indicating that the Weyl points move further away from the chemical potential as $\alpha$ increases.	
	
	\section{Experimental methods}
The synthesis of single-crystals of hexagonal-(Mn$_{1-\alpha}$Fe$_{\alpha}$)$_{3+\gamma}$Ge with varying Fe doping levels was carried out using a similar method described in Refs. \cite{rai_fe_neutron, rai_mn3ge}. The self-flux method was employed to synthesize the single-crystals under nearly identical conditions. Since the hexagonal phase of Mn$_3$Ge is stabilized with an excess of Mn ($\gamma$) \cite{berche2014thermodynamic, binary1}, hexagonal-(Mn$_{1-\alpha}$Fe$_{\alpha}$)$_{3+\gamma}$Ge compounds with $\gamma\approx0.2$ were synthesized.

The synthesis of (Mn$_{1-\alpha}$Fe$_{\alpha}$)$_{3.2}$Ge starts with the melting of  pure elements with a stoichiometric composition using the induction melting technique. The resulting samples were then sealed in quartz tubes and heated in the furnace up to 1273 K (for $\alpha=0$) or 1323 K (for $\alpha=0.30$) for 5 hours, followed by a slow cooling down to 1123 K, at the rate of 1 K/hr. It is important to mention that the Mn$_3$Ge and Fe$_3$Ge stabilize in the hexagonal crystal structure above 903 K and 973 K, respectively \cite{berche2014thermodynamic, binary1}. Therefore, all the samples were water-quenched at 1123 K to preserve the high-temperature hexagonal phase.

High-quality single-crystals were successfully obtained for compounds with Fe doping levels $\alpha=(0-0.22)$. However, the samples with $\alpha=0.26$ and $\alpha=0.30$ remained polycrystalline. The Laue diffraction patterns of the samples with $\alpha=0$ to $\alpha=0.22$ exhibited sharp 6-fold diffraction spots, indicating the formation of crystals in the hexagonal phase. An example of a Laue pattern for a selected sample with $\alpha=0.22$ is shown in Figure \ref{fig:laue} in the Appendix.

Chemical analysis of the crystals was performed using the ICP-OES (inductively coupled plasma optical emission spectroscopy) method. The analysis yielded the empirical formulas for samples with Fe doping levels $\alpha=0$, $\alpha=0.04$, $\alpha=0.10$, $\alpha=0.18$, $\alpha=0.22$, and $\alpha=0.30$ as Mn$_{3.10(5)}$Ge, (Mn$_{0.96(1)}$Fe$_{0.04(1)}$)$_{3.25}$Ge, (Mn$_{0.90(1)}$Fe$_{0.10(1)}$)$_{3.18}$Ge, (Mn$_{0.83(1)}$Fe$_{0.17(1)}$)$_{3.25}$Ge, (Mn$_{0.79(1)}$Fe$_{0.21(1)}$)$_{3.21}$Ge, and (Mn$_{0.68(1)}$Fe$_{0.32(1)}$)$_{3.20}$Ge, respectively. For simplicity, we will refer to these samples based on their initial stoichiometric composition.

	\begin{figure}
		\includegraphics[width=5.5cm]{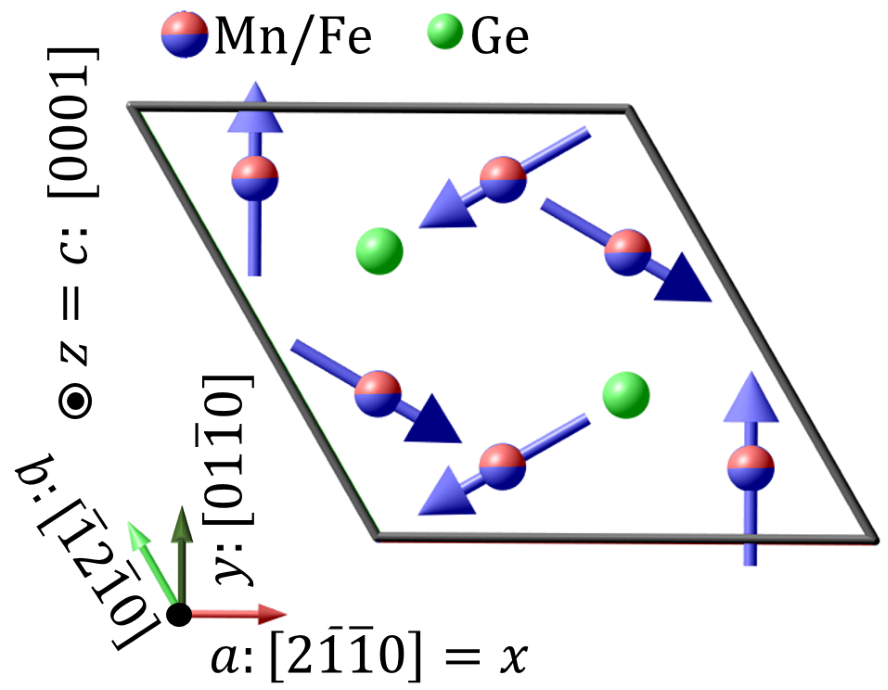} \caption{Magnetic structure of $\alpha=0.22$ compound at 130 K, as determined by the Ref. \cite{rai_fe_neutron}. The crystallographic axes \textit{a} and \textit{b}, related by 120\textdegree, show hexagonal coordinates.  The \textit{x, y,} and \textit{z} axes are described in the Cartesian coordinate system. The \textit{c} and \textit{z} axis remain the same in both coordinate systems.}
		\label{fig:nuclear}
	\end{figure}

A small amount of $\alpha=(0, 0.04, 0.10, 0.18, 0.22,$ $ 0.30)$ compounds were crushed and X-ray powder diffraction (XRPD) was performed using the Huber Imaging plate Guinier Camera (G670). The obtained data were analyzed using the FullProf software, and the results are presented in Figure \ref{fig:xrd_all}(a-f) in the Appendix. The analysis confirms that all the crystals were synthesized in the hexagonal phase with $P6_3/mmc$ (no. 194) space group symmetry. A small amount of tetragonal phase (2-4\%) was also observed as an impurity in the $\alpha=0$ and $\alpha=0.04$ compounds (Figure \ref{fig:xrd_all}(a, b) in the Appendix). This presence of a tetragonal phase has been commonly observed during the synthesis of hexagonal-Mn$_3$Ge \cite{chen2020antichiral, kiyohara2016giant}.

The lattice parameters of the samples in the hexagonal phase were compared and are shown in Figure \ref{fig:xrd_all}(g) in the Appendix. As expected, the lattice parameters and lattice volume decrease monotonically with an increase in Fe concentration, following Vegard's law \cite{vegardlaw}. Furthermore, neutron diffraction analysis of the $\alpha=0.22$  compound  \cite{rai_fe_neutron} confirmed the substitution of Mn by Fe on the Mn sites, while the excess Mn ($\gamma$) occupies the Ge sites. The nuclear and magnetic structure of the $\alpha=0.22$ compound at 130 K is illustrated in Figure \ref{fig:nuclear}. The plot also includes crystallographic axes in both hexagonal ($a, b$) and Cartesian ($x, y$) coordinates for convenience.
	
	\begin{figure} 
		\includegraphics[width=8cm]{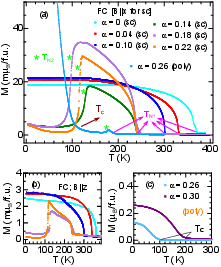} \caption{ The temperature-dependent magnetization (\textit{M-T}) measurements were performed under field-cooled (FC) conditions with a magnetic field of 0.01 T. The measurements shown here were taken during the warming of the sample. (a) Figure \ref{fig:MT_X_axis_Poly}(a) shows the \textit{M-T} curves for single-crystal (sc) (Mn$_{1-{\alpha}}$Fe$_{{\alpha}}$)$_{3.2}$Ge with $\alpha=0-0.22$ along the \textit{x} axis. The \textit{M-T} curve for the polycrystalline (poly) sample with $\alpha=0.26$ is also included for comparison. (b) Figure \ref{fig:MT_X_axis_Poly}(b) displays the \textit{M-T} curves for selected single-crystals with the magnetic field (\textit{B}) parallel to the \textit{z} axis. (c) Figure \ref{fig:MT_X_axis_Poly}(c) shows the \textit{M-T} curves for the (Mn$_{1-{\alpha}}$Fe$_{{\alpha}}$)$_{3.2}$Ge samples with $\alpha=0.26$ and $\alpha=0.30$.}
		\label{fig:MT_X_axis_Poly}
	\end{figure}

The magnetization and electrical transport measurements of all the samples were conducted using Quantum Design-Physical Property Measurement System (QD-PPMS) and Quantum Design-DynaCool (QD-DC) instruments. In the measurements, the magnetic and electrical properties were probed along the crystallographic axes $x$, $y$, and $z$ of the sample, as defined in Figure \ref{fig:nuclear}. This allows a comprehensive understanding of the anisotropic behavior and electronic transport characteristics of the samples.

The tetragonal and hexagonal phases of Mn$_3$Ge exhibit different magnetic behaviors, with the tetragonal phase being ferrimagnetic and the hexagonal phase being paramagnetic at 400 K. The Fe doped samples demonstrate paramagnetic behavior at 400 K, which is above the N\'{e}el temperature ($T_\text{N1}$) for $\alpha\leq0.26$ compounds. Therefore, it can be concluded that the observed transport and magnetic effects originate from the hexagonal phase of the compound. 

	\section{Magnetization} \label{magnetization}

	\begin{figure} 
		\includegraphics[width=8cm]{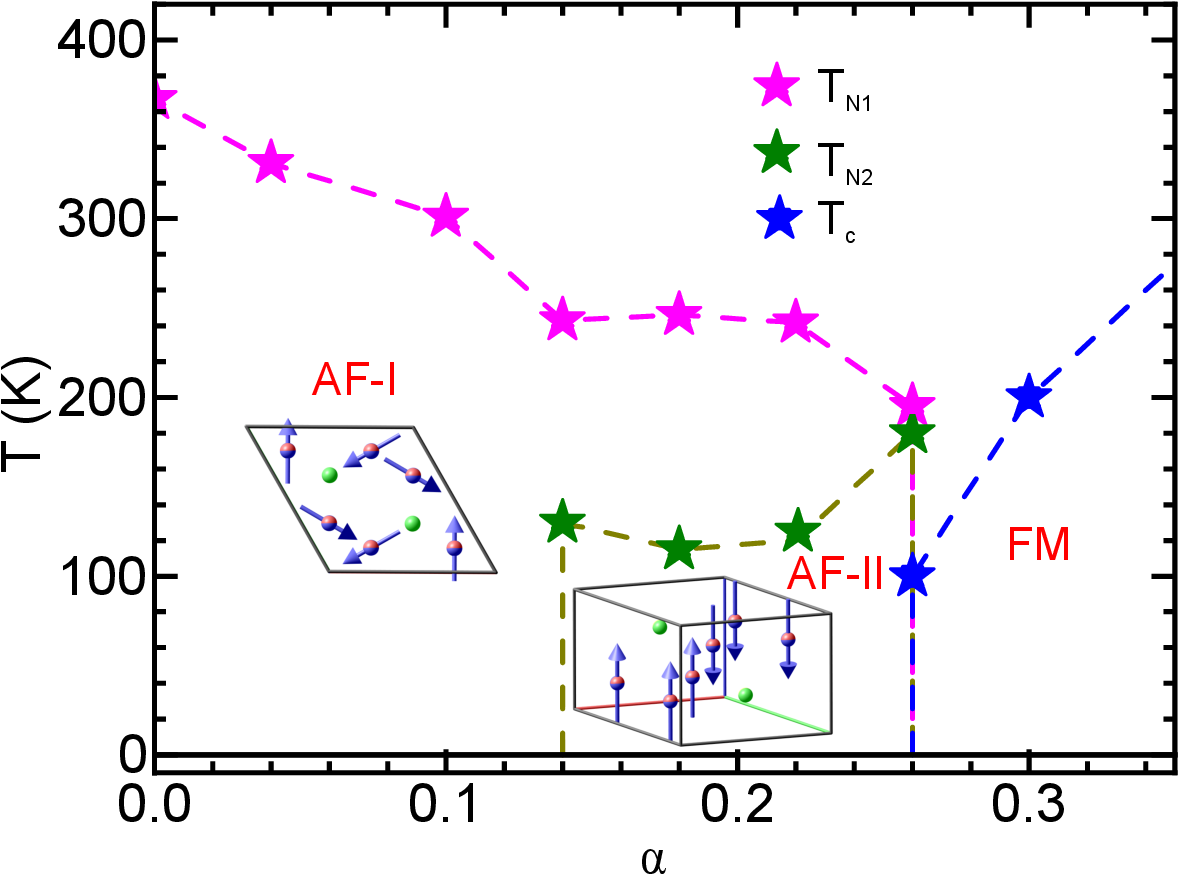} \caption{Magnetic phase diagram of the (Mn$_{1-{\alpha}}$Fe$_{{\alpha}}$)$_{3.2}$Ge ($\alpha=0-0.30$). $T_{\text{N1}}$, $T_{\text{N2}}$, and $T_{\text{c}}$ are the magnetic transition temperatures shown in Figure \ref{fig:MT_X_axis_Poly}. AF-I and AF-II show different antiferromagnetic regimes, whose corresponding magnetic structure is also shown in the same region. The samples with $\alpha>0.22$ show ferromagnetic (FM) behavior below $T_{\text{c}}$, as indicated in the figure.}
		\label{fig:phase_diagram}
	\end{figure} 
 
The temperature-dependent magnetization (\textit{M-T}) measurements were performed on single crystals and polycrystalline compounds with varying Fe doping levels ($\alpha=0-0.30$), and the results are shown in Figure \ref{fig:MT_X_axis_Poly}.

When measuring the magnetization along the \textit{x} axis, compounds with $\alpha=(0.04-0.26)$ exhibit a magnetic phase transition at the N\'{e}el temperature ($T_{\text{N1}}$), similar to Mn$_3$Ge \cite{rai_mn3ge} (Figure \ref{fig:MT_X_axis_Poly}(a)). Below $T_{\text{N1}}$, the magnetization along the \textit{x} axis starts to increase, resembling the behavior of the parent compound ($\alpha=0$). However, the compounds do not behave uniformly at low temperatures. In contrast to $\alpha\leq0.10$ compounds, compounds with $\alpha=(0.14-0.26)$ exhibit a second magnetic phase transition at $T_{\text{N2}}$ ($<T_{\text{N1}}$), as shown in Figure \ref{fig:MT_X_axis_Poly}(a). The magnetization magnitude below $T_{\text{N2}}$ remains significantly smaller compared to the magnetization between $T_{\text{N1}}$ and $T_{\text{N2}}$.

Magnetization measurements along the \textit{z} axis were also conducted for selected compounds ($\alpha=0, 0.04, 0.10, 0.18, 0.22$), as depicted in Figure \ref{fig:MT_X_axis_Poly}(b). The \textit{M-T} curve along the \textit{z} axis follows a similar pattern as that along the \textit{x} axis. However, the magnetization along the \textit{z} axis is nearly 10 times smaller than that along the \textit{x} axis throughout the measured temperature range.

The magnetization of $\alpha=0.26$ and $\alpha=0.30$ samples displays ferromagnetic (FM) behavior, and their respective Curie temperatures ($T_{\text{c}}$) are shown in Figure \ref{fig:MT_X_axis_Poly}(c).

The derivative of the $M-T$ curve was used to determine various magnetic transition temperatures. The evolution of $T_{\text{N1}}$, $T_{\text{N2}}$, and $T_{\text{c}}$ with Fe doping fraction ($\alpha$) is presented in the phase diagram shown in Figure \ref{fig:phase_diagram}. As mentioned earlier, the magnetization of $\alpha=(0.04-0.10)$ and $\alpha=(0.14-0.26)$ compounds exhibits behavior similar to that of Mn$_3$Ge below $T_{\text{N1}}$ and between $T_{\text{N2}}$ and $T_{\text{N1}}$, respectively. We define this magnetization regime as \textbf{AF-I}, as indicated in Figure \ref{fig:phase_diagram}. The magnetic structure of the $\alpha=0.22$ compound in the AF-I regime has already been reported to be the same as Mn$_3$Ge \cite{rai_fe_neutron}. Therefore, we can conclude that the magnetic structure of all compounds in the AF-I regime remains the same as that of the parent compound Mn$_3$Ge.

The $M(H)$ measurements for $\alpha=(0.14-0.26)$ compounds at 4 K reveal antiferromagnetic (AFM) behavior in the absence of residual magnetization. Previous studies \cite{hori1992antiferromagnetic, rai_fe_neutron} have reported that the $\alpha=0.22$ compound exhibits collinear AFM behavior with Mn moments aligned along the \textit{z} axis. Hence, it can be inferred that compounds with $\alpha=(0.14-0.26)$ possess a collinear AFM structure with Mn moments oriented along the \textit{z} axis below $T_{\text{N2}}$, as depicted in Figure \ref{fig:phase_diagram}. This magnetic region is referred to as the \textbf{AF-II regime} in the phase diagram. 
	
	\begin{figure} 
		\includegraphics[width=7.8cm]{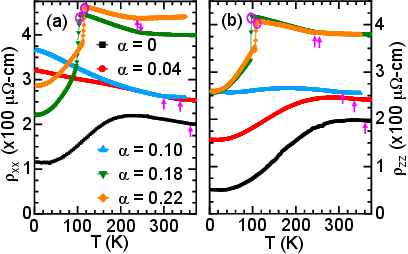} \caption{Temperature dependent longitudinal resistivity of the (Mn$_{1-{\alpha}}$Fe$_{{\alpha}}$)$_{3.2}$Ge ($\alpha=0-0.22$) along (a) \textit{x} axis, and (b) \textit{z} axis. Magenta arrows in (a, b) lie at $T_{\text{N1}}$, below which the respective samples lie in AF-I region.  Magenta circles in both the figures lie at $T_{\text{N2}}$, below which the respective samples lie in AF-II region.}
		\label{fig:rt_x_z_All}
	\end{figure}

	\section{Electrical Transport Results}
The electrical transport properties of single-crystal compounds with $\alpha=0, 0.04, 0.10, 0.18, 0.22$ were investigated. The crystal size $\alpha=0.14$ sample was very small. Therefore no electrical transport measurement was performed using this sample. The samples under measurements have typical dimensions of length $\simeq$ 1.0 mm - 1.5 mm, width $\simeq$ 0.4 mm - 0.5 mm, and thickness $\simeq$ 0.1 mm - 0.2 mm. Measurements were performed along the different crystallographic axes, and consistent results were obtained upon repeated measurements, confirming the intrinsic nature of the observed electrical transport effects.

The electrical resistivity of all the compounds was measured along the three different axes. We have observed similar resistivity  along the \textit{x} and \textit{y} axes, as reported by Refs. \cite{rai_fe_neutron, rai_mn3ge}. However, a significantly different resistivity behavior was observed along the \textit{z} axis, compared to the \textit{x} axis. The longitudinal resistivity along the \textit{x} and \textit{z} axes for $\alpha=0-0.22$ compounds is shown in Figure \ref{fig:rt_x_z_All}. The resistivity increases with increasing Fe doping fraction, which is expected due to impurity doping. The parent compound, $\alpha=0$, shows metallic behavior, along the \textit{x} axis, below 200 K. However, it shows semimetallic behavior above 200 K, up to $T_{\text{N1}}$. In contrast to this, all the Fe doped compounds exhibit semimetallic resistivity behavior along the \textit{x} axis in the entire AF-I regime. The resistivity along the \textit{z} axis shows metallic behavior for the parent compound. Whereas, it shows a mixture of metallic and semimetallic behavior for Fe doped compounds. The increase in resistivity, along both the axes, at low temperatures suggest strong spin scattering as magnetic moments of the compound increases. For $\alpha=(0.18, 0.22)$ compounds, the resistivity along both the \textit{x} and \textit{z} axes drops below $T_{\text{N2}}$ and exhibits metallic behavior in the AF-II regime.

	\subsection{Anomalous Hall effect}

	\begin{figure} [h]
		\includegraphics[width=8cm]{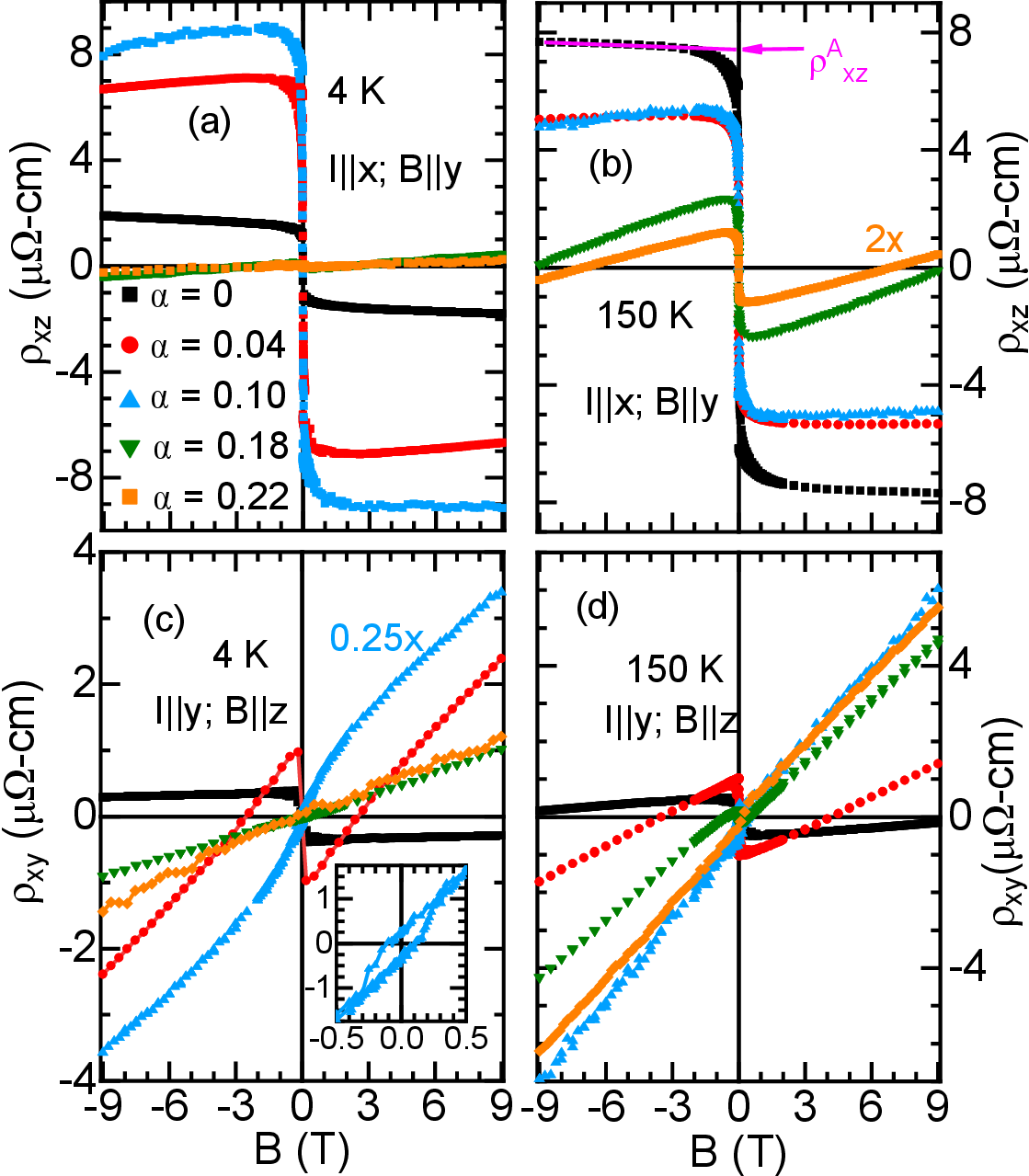} \caption{In Figures (a) and (b), the Hall resistivity measurements at 4 K and 150 K, respectively, are shown for the field applied along the \textit{y} axis. In Figures (c) and (d), the Hall resistivity measurements at 4 K and 150 K, respectively, are shown for the field applied along the \textit{z} axis. Inset of (c) represents the magnified version of the Hall resistivity ($\rho_{xy}$) for $\alpha=0.10$ compound. The Hall resistivity measurements were performed at 150 K, representing the AF-I regime for all the compounds. Hall resistivity at 4 K represents the AF-II regime for the $\alpha=0.18$ and $\alpha=0.22$ compounds only. In Figure (b) of the Hall resistivity measurements, $\rho_{xz}^A$ represents the anomalous Hall resistivity (AHR). It is determined by taking the intercept of the linear fit of the Hall resistivity data between -9 T and -3 T. } 
  
		\label{fig:hall_130_4k_All_doping}
	\end{figure}

	\begin{figure*}
		\includegraphics[width=17.0cm]{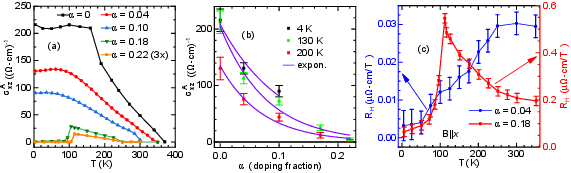} \caption{(a) The temperature dependence of the anomalous Hall conductivity (AHC) ($\sigma^A_{xz}$) is shown for the case where the current (\textit{I}) is parallel to the \textit{x} axis and the magnetic field (\textit{B}) is parallel to the \textit{y} axis. The data for $\alpha=0$ is obtained from Ref. \cite{rai_mn3ge}. (b) The evolution of the AHC ($\sigma^A_{xz}$) with respect to $\alpha$ (doping fraction) is presented at three different temperatures. At $T=4$ K, the AHC is zero for $\alpha=0.18$ and $\alpha=0.22$, and therefore not shown in this plot. The purple curves represent the exponential decay behavior (expon.) of the AHC. (c) The Hall coefficient ($R_H=\frac{\partial\rho_H}{\partial B}$) corresponding to the $\alpha=0.04$ and $\alpha=0.18$ compounds is shown, indicating their distinct magnetization behavior.}
		\label{fig:ahe_T_All doping}
	\end{figure*}

Hall resistivity measurements were performed on compounds with various Fe doping fractions, namely $\alpha=0, 0.04, 0.10, 0.18, 0.22$, with the magnetic field oriented along the \textit{y} and \textit{z} axes. Figure \ref{fig:hall_130_4k_All_doping}(a-d) displays the Hall resistivity values for these compounds at temperatures of 4 K and 150 K, with the magnetic field parallel to the \textit{y} and \textit{z} crystallographic axes. 

When the magnetic field was applied along the \textit{y} axis, a sharp jump in Hall resistivity was observed, in several cases, near zero fields, which  clearly suggests the presence of anomalous Hall resistivity (AHR).  The magnitude of the AHR was determined by performing a linear fit to the high-field Hall resistivity data, as shown in Figure \ref{fig:hall_130_4k_All_doping}(b). As depicted in Figure \ref{fig:hall_130_4k_All_doping}(a, b),  AHR ($\rho_{xz}^A$) is present at both 4 K and 150 K in compounds with Fe doping fractions of $\alpha=(0-0.10)$ and $\alpha=(0-0.22)$, respectively. Furthermore, the Hall hysteresis at 150 K, observed in both the parent and Fe-doped compounds, is confined to a range of 0.02 T, as illustrated in Figure \ref{fig:hall_systeresis} in the Appendix.

In comparison to the \textit{y} axis, when the magnetic field aligns parallel to the \textit{z} axis (\textit{B}$\parallel$\textit{z}), the observed Anomalous Hall Resistance (AHR) is much reduced, as depicted in Figure \ref{fig:hall_130_4k_All_doping}(c, d), or even absent. Prior research on Mn$_3$Ge has proposed that the AHR would vanish when the magnetic field aligns with the \textit{z} axis \cite{nayak2016large, rai_mn3ge}. However, we have observed a slight non-zero AHR for \textit{B}$\parallel$\textit{z}. Within this setup, the AHR magnitude is significantly prominent for $\alpha=0.04$, while it is minimal for $\alpha=0.10$ (shown in the inset of Fig. \ref{fig:hall_130_4k_All_doping}(c)). The presence of AHR for \textit{B}$\parallel$\textit{z} is not uniformly observed across all compounds. This inconsistency may stem from minor misalignment between the sample's orientation, the magnetic field's direction, and the crystallographic \textit{x} or \textit{y} axes.
Furthermore, the small out of the (\textit{a-b}) plane magnetization could also contribute to the minor AHR under the \textit{B}$\parallel$\textit{z} configuration \cite{kiyohara2016giant, rai_mn3ge,rai_fe_neutron, nayak2016large}. However, additional experimental investigations are necessary to precisely identify its underlying source.	
	
Multiple Hall resistivity ($\rho_{xz}$) measurements were conducted with a magnetic field applied along the \textit{y} axis at various temperatures. For each measurement, the corresponding AHR ($\rho^A_{xz}$) was determined using the linear fitting of the high field Hall resistivity data, as illustrated in Figure \ref{fig:hall_130_4k_All_doping}(b). In this context, $\rho^A_{xz}$ represents the AHR when the Hall voltage is measured along the \textit{z} axis, the current is applied along the \textit{x} axis, and the magnetic field is applied perpendicular to both axes, i.e. along the \textit{y} axis.

The Hall resistivity ($\rho_H$) can be expressed in a general form as $\rho_{\text{H}} = R_0B + R_MM+\rho_H^A$, where $R_0$ and $R_M$ correspond to the ordinary and magnetization induced Hall coefficients, respectively. $\rho_H^A$ denotes the anomalous Hall resistivity due to the non-vanishing Berry curvature \cite{kiyohara2016giant}. In compounds with ferromagnetic (FM) properties, the existence of AHR is expected due to the presence of remanent magnetization \cite{sales2006anomalous}. In such FM samples, the Hall resistivity exhibits a monotonic relationship with the magnetization of the sample under varying magnetic fields, as mentioned in Ref. \cite{sales2006anomalous}.

In the case of Fe-doped Mn$_3$Ge compounds, a small residual magnetization has been observed in the antiferromagnetic (AF-I) regime, indicating the presence of weak FM behavior (Figure \ref{fig:mh_All_T} in the Appendix). However, Refs. \cite{kiyohara2016giant, rai_fe_neutron} have reported that the Hall resistivity of compounds with $\alpha=(0, 0.22)$ in the AF-I regime does not exhibit a monotonic dependence on isothermal magnetization. This concludes that the observed AHR does not originate from the residual magnetization, but the non-vanishing Berry curvature, which leads to $\rho_H^A>0$. The nature of Hall resistivity of all the compounds between $\alpha=0$ and $\alpha=0.22$, within the AF-I regime, remains the same as long as the magnetic field is applied along the \textit{y} axis. Therefore, it can be concluded that the observed AHR (for \textit{B}$\parallel$\textit{y}), in the entire AF-I regime, originates from the non-vanishing Berry curvature \cite{kiyohara2016giant, nakatsuji2015large}. This also suggests the existence of Weyl points in the entire AF-I regime. 

The Fe doped Mn$_3$Ge compound with an $\alpha$ value of 0.30, which exhibits notable ferromagnetic behavior (refer to Figure \ref{fig:mh_All_T} in the Appendix), also demonstrates an AHR as depicted in Figure \ref{fig:lmr_30percent}(a) in the Appendix. By analyzing the relationship between AHR and magnetization of $\alpha=0.30$ compound, it is evident that an AHR of approximately 1.8 $\mu\Omega$ cm can be generated by a residual magnetization of 1.5 $\mu_{\textbf{B}}$/f.u.. In contrast to this, lower Fe doped compounds in the AF-I temperature range, show AHR $\geq0.8$ $\mu\Omega$ cm, even though their residual magnetization remains below 0.030 $\mu_{\textbf{B}}$/f.u.. Furthermore, examining the field-dependent magnetization (refer to Figure \ref{fig:mh_All_T} in the Appendix) of compounds with $\alpha=0-0.22$, we observe that the magnitude of residual magnetization of all compounds in the AF-I regime remains nearly equal at 150 K. However, the AHR at 150 K exhibits significant variations among all the compounds (refer to Figure \ref{fig:hall_130_4k_All_doping}(b)). These observations clearly indicate that the AHR observed in the AF-I regime does not originate from the residual magnetization in these compounds but the the non-vanishing Berry curvature caused by the existence of Weyl points within these compounds.

	The anomalous Hall conductivity (AHC), which is an intrinsic quantity, is proportional to the AHR of the sample. The AHC ($\sigma^A_{xz}$) can be determined using $\rho^A_{xz}$ using the relation:
	\begin{eqnarray}
		\sigma^A_{xz} \approx -\rho^A_{xz}/(\rho_{xx}\rho_{zz})
	\end{eqnarray}
	
The Hall resistivity for compounds with $\alpha=0-0.22$ was measured at various temperatures, and the corresponding AHR was calculated. Using the AHR values at different temperatures and the relation mentioned above, the temperature dependence of the AHC ($\sigma^A_{xz}$) was determined for these compounds, as depicted in Figure \ref{fig:ahe_T_All doping}(a). It is noteworthy that the AHC remains non-zero throughout the AF-I regime and becomes zero for $\alpha=0.18$ and $\alpha=0.22$ below 110 K, which corresponds to the AF-II regime. This observation indicates the existence of Weyl points in the entire AF-I regime, which disappears in the AF-II regime. The vanishing AHC in the AF-II regime is expected due to the underlying magnetic symmetry, as explained in Ref. \cite{rai_fe_neutron}, suggesting a strong connection between the Weyl points and the magnetic symmetry of the system.

In Figure \ref{fig:ahe_T_All doping}(b), it is evident that the anomalous Hall conductivity (AHC) at 130 K and 200 K (in the AF-I regime) exhibits an exponential decay pattern as the Fe doping fraction ($\alpha$) increases. The significant decrease in AHC with small Fe doping indicates that the Weyl points can be substantially modified by suitable dopants in a Weyl semimetal. Similar decrease in AHC with increasing Mn concentration have also been reported for Mn$_3$Ge and Mn$_3$Sn compounds \cite{chen2021anomalous, kiyohara2016giant}.
The strength of the Berry curvature in an ideal Weyl semimetal, where the Weyl points are separated by $\Delta k$ in phase space, can be described by the following equation 
\cite{vsmejkal2018topological, burkov2014anomalous, soh2019ideal}
	\begin{eqnarray}
		\sigma^A_{ij}=\frac{e^2}{2\pi h}\Delta k
	\end{eqnarray}
 
This implies that with an increase in Fe doping, the separation between a pair of Weyl points decreases \cite{vsmejkal2018topological, burkov2014chiral}. However, since the anomalous Hall effect (AHE) is also influenced by the relative position of the Weyl points with respect to the Fermi surface {\cite{chen2021anomalous}}, it is possible that the Weyl points move away from the Fermi surface as Fe doping fraction increases in Mn$_3$Ge. Similar decrease in AHE due to increase in separation between Weyl points and Fermi surface has been reported by Refs. \mbox{\cite{chen2014anomalous, kiyohara2016giant}} as well. Therefore, it can be concluded that the properties of Weyl points in Mn$_3$Ge can be adjusted by introducing Fe dopants. To accurately determine the precise changes in the separation between Weyl points and their relative position to the Fermi surface, a comprehensive theoretical calculation is required.

The Hall coefficient ($R_H$) can be determined by calculating the slope of the Hall resistivity at high magnetic fields: $R_H=\frac{\partial\rho_H}{\partial B}$. The values of $R_H$ for $\alpha=0.04$ and $\alpha=0.18$ were calculated and are shown in Figure \ref{fig:ahe_T_All doping}(c). Interestingly, unlike Mn$_3$Ge \cite{rai_mn3ge}, the sign change in $R_H$ does not occur below $T{_\text{N1}}$ (325 K) for $\alpha=0.04$ down to 4 K. Furthermore, Figure \ref{fig:ahe_T_All doping}(c) demonstrates that $R_H$ for $\alpha=0.18$ exhibits a sudden drop near 110 K, similar to the behavior observed in temperature dependent magnetization and resistivity of the same sample. This suggests a significant increase in carrier concentration below $T_{_\text{N2}}$ as the Mn spins flip towards the \textit{z} axis.

	\begin{figure}[h]
		\includegraphics[width=8.4cm]{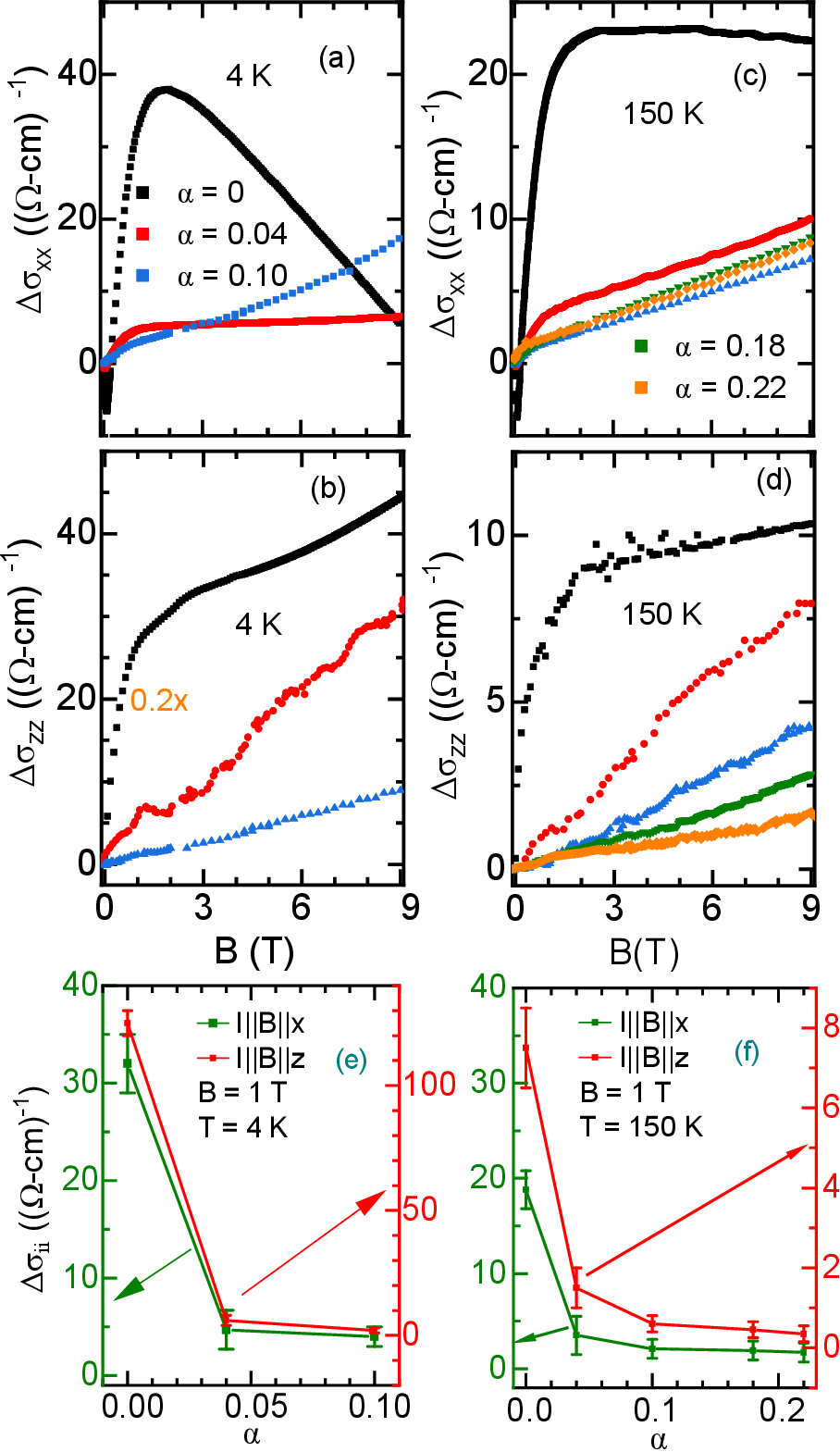} \caption{LMC of single-crystal samples with $\alpha=0-0.22$. (a - d) LMC along the \textit{x} and \textit{z} axis at 4 K and 150 K. (e, f) Variation of the LMC (at 1 T) with the Fe doping fraction ($\alpha$).}
		\label{fig:lmc-B_All}
	\end{figure} 
	
	\subsection{Magnetoconductance and planar Hall effect}
 
 The observation of AHE in the coplanar triangular antiferromagnetic (AF-I) regime underpins the existence of Weyl points in this regime. In addition to AHE, the chiral anomaly is also a well-known phenomenon in Weyl semimetals. Therefore, to identify the characteristic signatures of the chiral anomaly effect, we performed longitudinal magnetoconductivity, angular magnetoconductivity, and planar Hall effect measurements on the Fe-doped Mn$_3$Ge compounds. We will discuss each of these measurements in detail below. 
	
	\subsubsection{Longitudinal magnetoconductivity (LMC)} \label{LMC}
	The longitudinal magnetoresistivity (LMR) of compounds with different Fe doping levels, represented by $\alpha=$ (0, 0.04, 0.10, 0.18, 0.22), was measured over a temperature range starting above $T_{\text{N1}}$ and down to 4 K. LMR along the $i$ axis ($i= x z$ in our case) is denoted as $\Delta\rho_{ii}$, and defined as: $\Delta\rho_{ii}=\rho_{ii}(B)-\rho_{ii}(0)$, where $\rho_{ii}$ denotes the longitudinal resistivity. By utilizing $\Delta\rho_{ii}$ and the conductivity $\sigma_{ii}=1/\rho_{ii}$, the longitudinal magnetoconductivity (LMC), denoted as $\Delta\sigma_{ii}$, can be calculated using the following relationship:
	\begin{equation}				\Delta\sigma_{ii}=\sigma_{ii}(B)-\sigma_{ii}(0)\approx \frac{-\Delta\rho_{ii}}{\rho^2_{ii}}		
	\end{equation}
	
LMC for various compounds, at different temperatures, was determined and shown in Figure {\ref{fig:lmc-B_All}}. The analysis of LMC is focused at  4 K and 150 K. For compounds with $\alpha\leq0.10$, both temperatures lie in the AF-I regime. However, for compounds with $\alpha=(0.18, 0.22)$, 4 K and 150 K correspond to the AF-II and AF-I regimes, respectively. Since the presence of Weyl points is expected only in the AF-I regime, the discussion of chiral anomaly-induced effects will be limited to the samples and temperatures which belong to the AF-I regime. The LMC in the AF-II regime ($\alpha=0.18, 0.22$; $T=4$ K), where chiral anomaly-induced effects are not expected, is addressed in the Appendix \ref{appendix:electrical}. 

Positive LMC was observed in compounds with $\alpha=(0-0.22)$, particularly when a low magnetic field is applied in the AF-I regime. However, the origin of this positive LMC is not easily discernible since it can arise from various effects. The most likely factors contributing to the observed positive LMC include the current jetting effect \cite{liang2018experimental}, magnetization \cite{ritchie2003magnetic}, the unequal spin density of states (sDOS) in halfmetallic/semimetallic compounds \cite{yang2012anisotropic, roth1963empirical, furukawa1962magnetoresistance, kawabata1980theory, Endo1999MagnetoresistanceOC, kokado2012anisotropic}, and the chiral anomaly effect \cite{burkov2016z}. Each of these possibilities will be discussed in detail below.

	\begin{figure*} 
		\includegraphics[width=16cm]{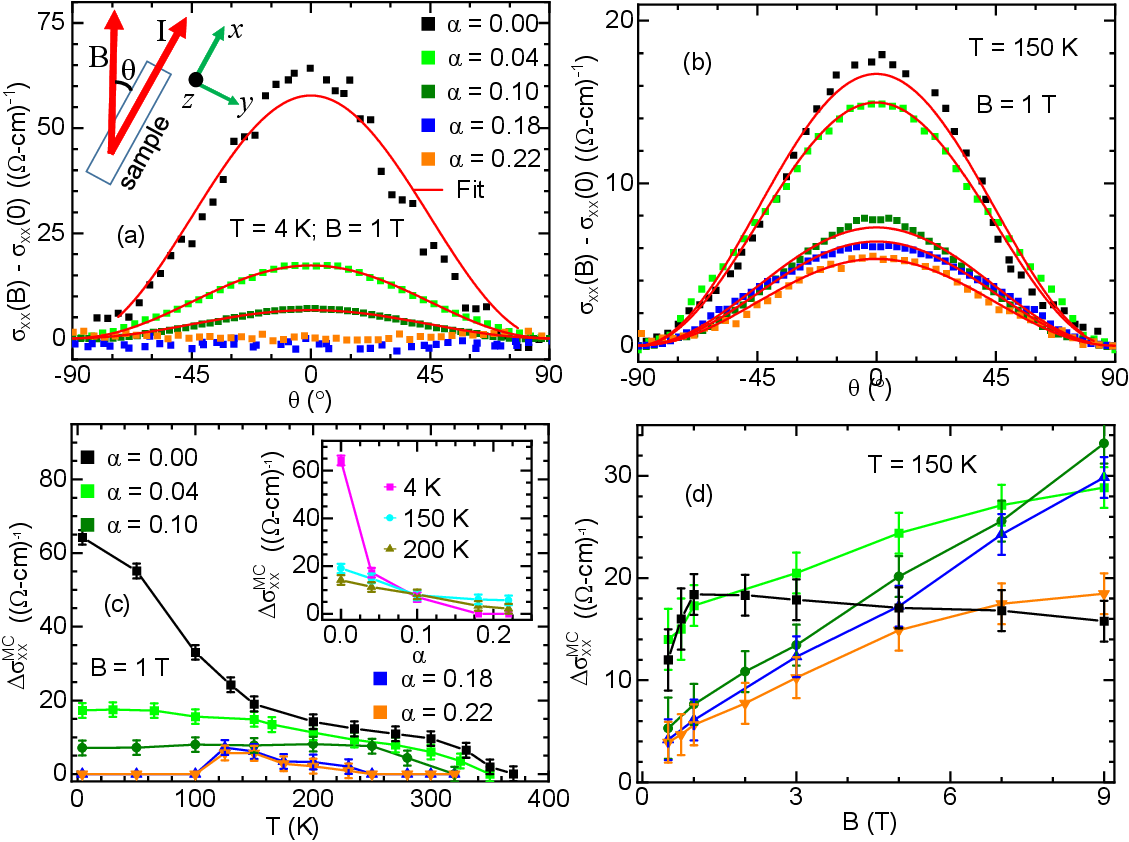} \caption{$\theta$MC for $\alpha=0-0.22$ compounds with sample rotated from \textit{x} axis, towards the \textit{y} axis as shown in (a). $\theta$MC for $\alpha=(0, 0.04, 0.10, 0.18, 0.22)$, measured at 1 T, is compared at 4 K and 150 K in (a), and (b) respectively. (c) Evolution of magnitude of $\theta$MC ($\Delta\sigma_{xx}^{\text{MC}}$) with temperature and  magnetic field. Here, $\Delta\sigma_{xx}^{\text{MC}}$ signifies the magnitude of oscillation determined by $\sigma(0$\textdegree$)-\sigma(90$\textdegree$)$, keeping the magnetic field and temperature constant. (c) Inset: Evolution of $\theta$MC magnitude with Fe doping fraction ($\alpha$), compared at 4 K, 150 K, and 200 K. (d) Evolution of the magnitude of $\theta$MC with the magnetic field.}
		\label{fig:mr_th_all}
	\end{figure*}

{\textit{Current jetting effect}}: The influence of the current jetting effect was investigated in both the AF-I and AF-II regimes of Fe-doped compounds, as described in Figure {\ref{fig:current_jetting}} (Appendix {\ref{appendix:electrical}}). Some of the data shown in Figure {\ref{fig:lmc-B_All}} (AF-I regime) is shown (in raw form) in Figure {\ref{fig:current_jetting}} as well, to determine the effect of current jetting in the AF-I regime.  The raw data for $\alpha=0.18$ at 4 K, which belongs to the AF-II regime, is also shown in the same figure to determine the effect of current jetting in the AF-II regime. Our measurements on samples with $\alpha=0.10$ and $\alpha=0.18$ did not reveal a significant contribution of the current jetting effect to the observed LMR in AF-I and AF-II regime.  Consistent results were obtained from repeated measurements using different sample pieces. These findings suggest that the current jetting effect plays a negligible role in Fe-doped compounds, and the observed electrical transport effects are intrinsic. Similar observation has been reported for Mn$_3$Ge as well {\cite{rai_mn3ge}}.

\textit{Magnetization}: The role of magnetization in the LMC of Mn$_3$Ge has already been dismissed in Ref. \cite{rai_mn3ge}, as no correlation between the magnitude of LMC and sample magnetization was observed. In Fe-doped compounds, the magnetization in the AF-I regime remains similar and greater in magnitude compared to Mn$_3$Ge. However, the magnitude of LMC decreases significantly even with a small Fe doping fraction of 4\% in Mn$_3$Ge. Furthermore, the magnitude of LMC at 1 T continues to decrease significantly with an increase in Fe doping fraction (Figure \ref{fig:lmc-B_All}). Therefore, the role of magnetization in the observed LMC of Fe-doped Mn$_3$Ge compounds within the AF-I regime is considered negligible.
	
	\textit{Unequal sDOS in semi/half-metallic compounds}: According to Ref. {\cite{rai_mn3ge}}, it is important to consider the possibility of trivial magnetoconductance due to unequal spin density of states (sDOS), near the Fermi surface, in semi/half-metallic compounds, which lead to the unequal spin scattering, and may result in positive LMC \cite{yang2012anisotropic, roth1963empirical, furukawa1962magnetoresistance, kawabata1980theory, Endo1999MagnetoresistanceOC, kokado2012anisotropic, bhattacharyya2019spin}. In the case of the Fe-doped compounds with doping fractions $\alpha=0.04-0.22$ (Figure \ref{fig:rt_x_z_All}), it is noteworthy that the resistivity along the \textit{x} axis is semi/half-metallic within the AF-I regime. Additionally, compounds with $\alpha=0.10-0.22$ exhibit semi/half-metallic behavior along the \textit{z} axis as well. The semi/half-metallic nature of each compound can be confirmed through spin density of states (sDOS) calculations.

Typically, metallic compounds exhibit negative LMC, while semi/half-metallic compounds show positive LMC \cite{kokado2012anisotropic}. In the case of the Fe-doped compounds, they exhibit semi/half-metallic resistivity along the \textit{x} axis in the AF-I regime, which justifies the observed positive LMC (Figure \ref{fig:lmc-B_All}(a, c)). However, compounds with $\alpha=0$ and $\alpha=0.04$, which are metallic along the \textit{z} axis, also display positive LMC along the same axis at both 4 K and 150 K (Figure \ref{fig:lmc-B_All}(b, d)). Consequently, the role of unequal sDOS alone cannot satisfactorily explain the observed positive LMC within the AF-I regime. Therefore, it can be concluded that the role of unequal sDOS in explaining the observed positive LMC within the AF-I regime is not sufficient and requires further investigation and consideration of other factors.

	\textit{Chiral anomaly effect}: Figure \ref{fig:lmc-B_All}(a-d) provides clear evidence of a positive LMC signature along both the \textit{x} and \textit{z} axes in all Fe-doped compounds within the AF-I regime. The decrease in LMC along the \textit{x} axis for $\alpha=0$, beyond 1.5 T at 4 K and 150 K, can be attributed to the semi/half-metallic nature of Mn$_3$Ge \cite{rai_mn3ge}. However, this decrease is not observed beyond 4\% Fe doping, indicating that impurities can affect the spin density of states (sDOS) near the Fermi surface, which typically leads to a negative slope in LMC for metallic systems.
	
	\begin{figure*} 
		\includegraphics[width=16cm]{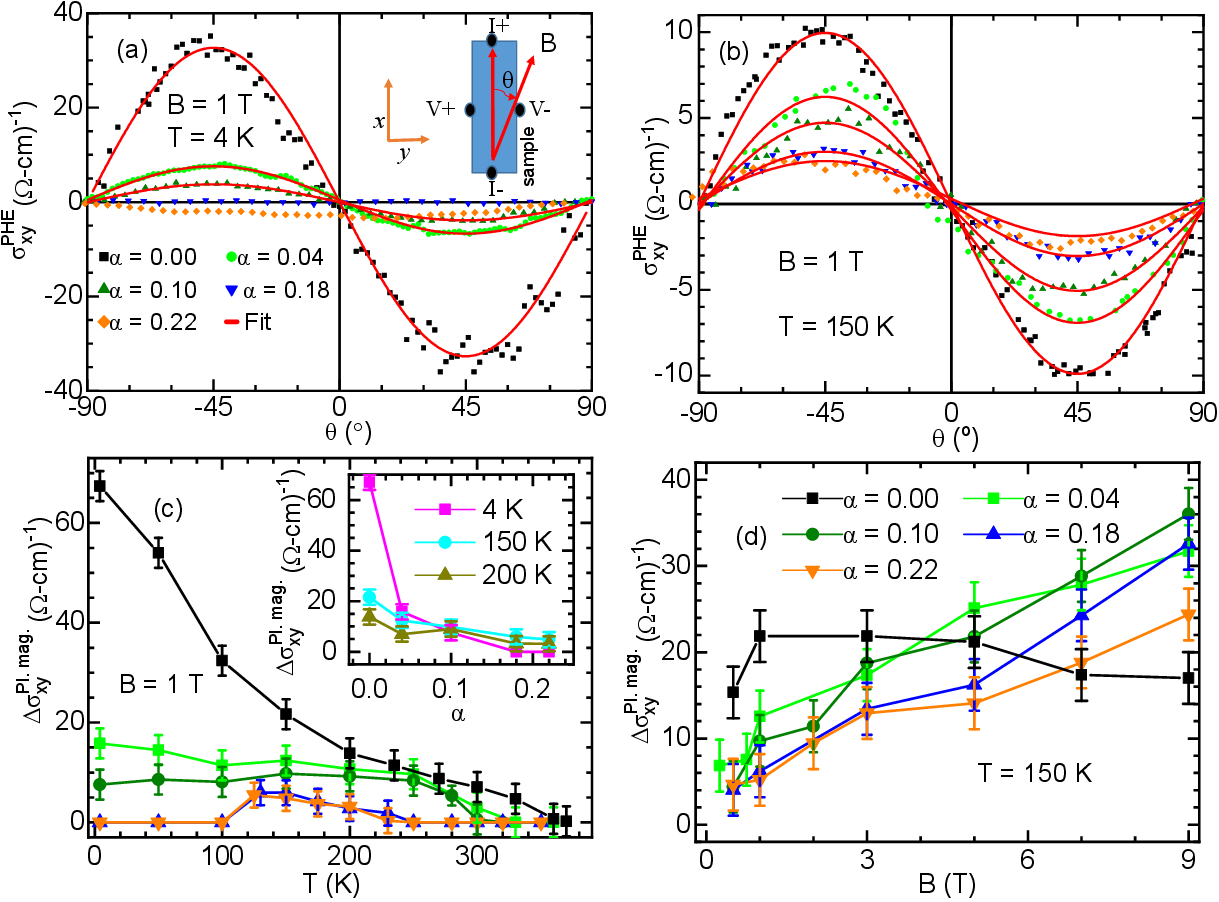} \caption{PHE for different compounds ($\alpha=0-0.22$) at 4 K and 150 K is shown in (a) and (b), respectively. The measurements were performed at 1 T of the applied magnetic field. The setup for the PHE measurement in the $x-y$ plane is illustrated in (a). (c, d) Variation of PHE magnitude ($\Delta\sigma_{xx}^{\text{Pl. mag.}}$) with temperature and magnetic field. In this case, $\Delta\sigma_{xx}^{\text{Pl. mag.}}$ is determined by the data fitting using the Eqn. \ref{eq:phe}. (c) Inset: Evolution of magnitude of PHE oscillation ($\Delta\sigma_{xx}^{\text{Pl. mag.}}$) with $\alpha$ (doping fraction).}
		\label{fig:phe_all}
	\end{figure*}

Positive LMC induced by the chiral anomaly has been reported previously in Mn$_3$Ge and Mn$_3$Sn \cite{chen2021anomalous, kuroda2017evidence, wu2023temperature}. Given that the doped compounds in the AF-I regime also exhibit positive and increasing LMC, it is plausible to suggest that the origin of the positive LMC within the range of $\alpha=0-0.22$ in the AF-I regime is the chiral anomaly effect.

Figure \ref{fig:lmc-B_All}(e, f) presents a comparison of the  LMC evolution at 1 T with varying Fe doping fractions ($\alpha$). It is evident that the LMC along the \textit{x} and \textit{z} axes experience a sharp decrease as Fe doping increases at 4 K. Similarly, even with a 4\% Fe doping, the LMC exhibits a significant reduction at 150 K. This abrupt decline in positive LMC, despite minor impurity doping, can be attributed to the chiral anomaly-induced LMC phenomenon. Previous studies \cite{kuroda2017evidence, chen2021anomalous} have suggested that the position of Weyl points relative to the chemical potential ($\mu$) is sensitive to impurities. Furthermore, according to Refs. \cite{burkov2016z, burkov2017giant}, the chiral anomaly-induced LMC ($\Delta\sigma_{\text{chiral}}$) is inversely proportional to $\mu^2$, expressed as $\Delta\sigma_{\text{chiral}}\propto\frac{1}{\mu^2}$. Thus, the sudden decrease in LMC at 1 T following a small amount of Fe doping implies that the LMC in doped compounds may originate from the chiral anomaly effect.

It is worth noting that positive LMC is also observed in the AF-II regime, at low field, as depicted in Figure \ref{fig:LMC_AF2} in the Appendix. However, the chiral anomaly-induced LMC is not expected to be present in this region because of the lack of evidence of Weyl points in the AF-II regime \cite{rai_fe_neutron}. Therefore, the presence of the chiral anomaly effect cannot be solely justified based on the positive LMC observed in the AF-I regime, and further investigations are required to justify the evidence of the chiral anomaly effect in Fe doped compounds.

	\subsubsection{Angular magnetoconductivity ($\theta$MC)}\label{AMC}
The detection of anisotropy in angular magnetoconductivity ($\theta$MC) measurements is a prominent indication of the chiral anomaly effect, as discussed in Refs. \cite{burkov2014chiral} and \cite{burkov2017giant}. Therefore, we carried out $\theta$MC measurements on compounds doped with $\alpha=0-0.22$. As mentioned in 
Refs. \cite{rai_fe_neutron, rai_mn3ge, nayak2016large}, the magnetization along the \textit{x} and \textit{y} axes remains nearly the same for parent and 22\% Fe doped Mn$_3$Ge compounds, whereas, it differs significantly along the \textit{z} axes (section \ref{magnetization}). Along with the chiral anomaly effect, the anisotropy in magnetization within a particular plane can also lead to the observation of anisotropic magnetoconductivity if measured within the same plane. Consequently, to eliminate the contribution of magnetoconductivity arising from the difference in magnetization along the \textit{x} and \textit{z} axes, we performed the $\theta$MC measurements exclusively within the \textit{x-y} plane. This choice was made due to the similar magnitude and behavior of magnetization observed along the \textit{x} and \textit{y} axes.

The setup for the $\theta$MC measurement is illustrated in Figure \ref{fig:mr_th_all}(a), where the longitudinal magnetoconductivity of the sample was measured along the \textit{x} axis while rotating the sample in the \textit{x-y} crystallographic plane. The \textit{x} axis of the sample was oriented at an angle $\theta$ with respect to the applied magnetic field. Here, $\theta=0$ represents the longitudinal magnetoconductance (LMR), while $\theta=90^\circ$ corresponds to the magnetoconductance where the applied magnetic field is perpendicular to the electric current direction.
	
The $\theta$MC measurements were performed on the compounds with Fe doping fractions $\alpha=0-0.22$ at constant temperatures and magnetic fields. Figure \ref{fig:mr_th_all}(a, b) shows the observed angular anisotropy in $\theta$MC for most of the compounds. The $\theta$MC values at 1 T for different compounds are compared at 4 K and 150 K. At 4 K, angular anisotropy in $\theta$MC was observed only for $\alpha=0, 0.04, 0.10$ compounds. In contrast, at 150 K, significant angular anisotropy in $\theta$MC was observed for all the compounds within the range of $\alpha=0-0.22$.

For compounds with a small LMC $(\Delta\sigma_{xx}(B)=(\sigma_{xx}(B)-\sigma_{xx}(0))<<\sigma_{xx}(0)))$, the chiral anomaly-induced $\theta$MC can be described by the following equation \cite{deng2020_chiral, burkov2014chiral, burkov2017giant}:

\begin{eqnarray}
		\sigma_{xx}(\theta)-\sigma_{xx}(\perp)\approx \Delta\sigma_{xx}^{\text{MC}}\text{cos}^2\theta\nonumber\\
		\Delta\sigma_{xx}^{\text{MC}}=(\sigma_{xx}^{\text{MC}}(\parallel)-\sigma_{xx}^{\text{MC}}(\perp)) 
		\label{eqn:mc}
	\end{eqnarray}

Where, $\Delta\sigma_{xx}^{\text{MC}}$ correspond to the magnitude of angular variation of $\theta$MC. Ideally, at constant temperature, $\Delta\sigma_{xx}^{\text{MC}}(B)= (\sigma_{xx}(B)_{\theta=0^o}-\sigma_{xx}(0)_{\theta=0^o})$. However, it may differ due to the presence of a transverse magnetoconductivity, 
which is generally non-zero in most cases.

From Figure \ref{fig:mr_th_all}(a, b), it can be observed that the $\theta$MC for almost all the compounds fits well with Equation \ref{eqn:mc}. Using this fitting, $\Delta\sigma_{xx}^{\text{MC}}$ was determined for all the compounds at various temperatures and compared in Figure \ref{fig:mr_th_all}(c). Notably, the non-zero magnitude of $\theta$MC is observed only in the temperature and doping regime where the magnetic structure is similar to Mn$_3$Ge, i.e., in the AF-I regime. $\theta$MC becomes negligible for $\alpha=(0.18, 0.22)$ below approximately 110 K, which is close to the $T_{\text{N2}}$ of the corresponding compound.

The evolution of $\Delta\sigma_{xx}^{\text{MC}}$, at constant temperatures, with Fe doping fraction is also shown in the inset of Figure \ref{fig:mr_th_all}(c), revealing a significant decrease in $\Delta\sigma_{xx}^{\text{MC}}$ with Fe doping, similar to the LMC of the same compound (Figure \ref{fig:lmc-B_All}(e, f)). This suggests that the origin of angular anisotropy in $\theta$MC is likely to be the chiral anomaly effect, which is possibly responsible for the positive LMC observed in these compounds in the AF-I regime. Additionally, it was observed that $\Delta\sigma_{xx}^{\text{MC}}$ for $\alpha\geq0.04$ increases almost linearly with the increase in magnetic field (Figure \ref{fig:mr_th_all}(d)), which is expected in the case of chiral anomaly effects in type-II Weyl semimetals \cite{nandy2017chiral}. This suggests that the Fe-doped compounds may also host type-II Weyl points, similar to Mn$_3$Ge, within the AF-I regime \cite{yang2017topological}. The decrease in $\Delta\sigma_{xx}^{\text{MC}}$ with magnetic field for $\alpha=0$ has been explained in Ref. \cite{rai_mn3ge}.

	\subsubsection{Planar Hall Effect (PHE)}
The Planar Hall conductivity, also known as the Planar Hall effect (PHE), was measured for the compounds with Fe doping fractions $\alpha=0-0.22$ by rotating the magnetic field within the \textit{x-y} plane, as depicted in Figure \ref{fig:phe_all}(a). The measurements were performed within this plane for the same reasons mentioned earlier for the $\theta$MC measurements.

PHE measurements were conducted at 4 K and 150 K with an applied magnetic field of 1 T. The PHE data for various compounds are presented in Figure \ref{fig:phe_all}(a, b). Similar to the $\theta$MC results, PHE oscillations with a 180$^\circ$ periodicity are observed at 4 K for $\alpha=0-0.10$, but they disappear for higher doping fractions. However, at 150 K, PHE oscillations are observed for all the compounds within the measured range of $\alpha=(0-0.22)$.

The angular dependence of chiral anomaly-induced PHE, at a given magnetic field and temperature, follows the relation \cite{burkov2017giant}:
	
	\begin{eqnarray}
		\sigma_{xy}^{\text{PHE}}(\theta) \approx \Delta\sigma_{xx}^{\text{Pl. mag.}}[\sin\theta\cos\theta]
		\label{eq:phe}
	\end{eqnarray}
	
Here, $\Delta\sigma_{xx}^{\text{Pl. mag.}}$ represents twice the magnitude of the PHE oscillations. Ideally, at a constant temperature, $\Delta\sigma_{xx}^{\text{Pl. mag.}}(B)=(\sigma_{xx}(B)_{\theta=0}-\sigma_{xx}(0)_{\theta=0})$, similar to the magnitude of the $\theta$MC.

As shown in Figure \ref{fig:phe_all}(a, b), the PHE data fit reasonably well with Equation \ref{eq:phe}. Furthermore, the evolution of $\Delta\sigma_{xx}^{\text{Pl. mag.}}$ for all the doped compounds at 1 T is analyzed in Figure \ref{fig:phe_all}(c). It is interesting to observe that $\Delta\sigma_{xx}^{\text{Pl. mag.}}$ varies with temperature and doping fraction ($\alpha$) in a similar manner as observed in the case of $\theta$MC (compare Figure \ref{fig:mr_th_all}(c) and Figure \ref{fig:phe_all}(c)). Once again, it is evident that the PHE is present only in the AF-I regime. Additionally, the PHE significantly weakens with an increase in Fe doping, similar to the observations in LMC and $\theta$MC. Moreover, as depicted in Figure \ref{fig:phe_all}(d), $\Delta\sigma_{xx}^{\text{Pl. mag.}}$ shows a roughly linear increase with the magnetic field at 150 K, which aligns with the field dependence of $\theta$MC (Figure \ref{fig:mr_th_all}(d)).
	
	\textbf{Analysis}: The similarity in the magnitude of the PHE $(\Delta\sigma_{xx}^{\text{Pl. mag.}}(B, T, \alpha))$ and the magnitude of the LMC $(\Delta\sigma_{xx}^{\text{MC}}(B, T, \alpha))$ can be attributed to their common origin from the chiral anomaly effect \cite{deng2020_chiral, nandy2017chiral, burkov2017giant, burkov2016z, burkov2014chiral}. The observation of both $\theta$MC and PHE only within the AF-I regime suggests that their possible origin is the chiral anomaly effect. The significant decrease in the magnitude of $\theta$MC, PHE, and LMC with Fe doping further supports the role of the chiral anomaly in the observed behavior of $\theta$MC and PHE. This weakening of the effects also implies that the Weyl points move significantly further from the Fermi surface as the Fe doping fraction increases in Mn$_3$Ge.

	\section{Conclusion}
Extensive studies were conducted on the magnetization behavior of both the parent Mn$_3$Ge compound and Fe-doped Mn$_3$Ge compounds. The results showed that compounds with Fe doping fractions up to $\alpha=0.26$ exhibit magnetization and magnetic structure similar to Mn$_3$Ge, within the AF-I temperature regime. Notably, intermediate Fe-doped compounds displayed magnetization behavior that differed significantly from that of Mn$_3$Ge.

Further investigations focused on the electrical transport properties of compounds with Fe doping fractions ranging from $\alpha=0$ to $\alpha=0.22$. The AHE was found exclusively within the AF-I regime for all these compounds. Signatures of Weyl points were observed in a significantly large fraction of Fe doping, which highlights the robust and topologically protected nature of Weyl points in doped compounds. The AHE weakened considerably with increasing Fe doping concentration, indicating changes in the separation of Weyl point pairs and the position of Weyl points relative to the chemical potential as the Fe doping levels increased \cite{son2013chiral, vsmejkal2018topological}. 

Furthermore, positive LMC, non-zero $\theta$MC, and PHE were observed in the AF-I regime of the doped compounds. Analysis suggested that these effects likely originated from the chiral anomaly phenomenon. However, additional experimental and theoretical investigations are necessary to precisely determine the contributions of the chiral anomaly and other effects in the observed electrical transport features within the AF-I regime. The weakening of LMC, $\theta$MC, and PHE with increasing Fe doping implied that the Weyl points moved further away from the chemical potential as the Fe doping fraction in Mn$_3$Ge increased {\cite{son2013chiral}}. In conclusion, appropriate doping in the parent Weyl semimetallic system can significantly control the characteristics of Weyl points.

It is noteworthy that the AHE and all other effects, which can be associated with the chiral anomaly, completely vanish in the AF-II regime, which corresponds to a collinear antiferromagnetic magnetic structure, as shown in the phase diagram. This clear absence of Weyl points in the AF-II regime indicates that the existence of Weyl points in Fe-doped Mn$_3$Ge compounds, as well as similar compounds, is dictated by the magnetic symmetry of the compound.

	\global\long\def\appendixname{APPENDIX}%
	\appendix

	\section{SAMPLE CHARACTERIZATION}
	\subsection*{X-ray diffraction and magnetization} \label{appendix:exp}

	\begin{figure}[h]
		\includegraphics[width=7.4cm]{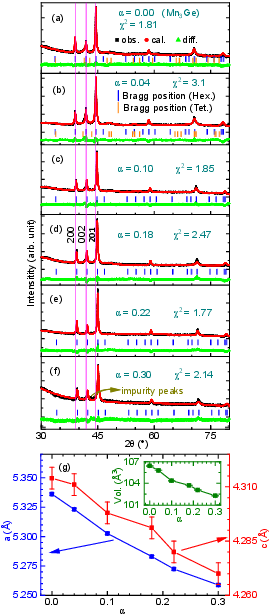} \caption{(a-f) Room temperature X-ray powder diffraction (XRPD) of $\alpha=0-0.30$ compounds. The arrows signify unknown impurity peaks. The goodness of the fitting parameter, $\chi^2$, for each compound is mentioned in each plot. Obs., cal., implies the observed and calculated XRPD pattern. Diff. shows the difference between observed and calculated intensity. Hex. and Tet. denote hexagonal and tetragonal phases, respectively. (g) Variation of lattice parameters with the doping fraction ($\alpha$). Inset: Variation of the lattice volume with $\alpha$.}
		\label{fig:xrd_all}
	\end{figure}
	
	\begin{figure}[h]
		\includegraphics[width=7.5cm]{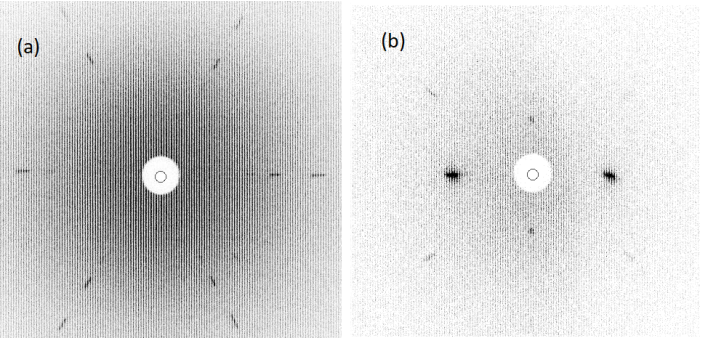} \caption{Laue diffraction pattern corresponding to the X-ray beam along the (a) \textit{z} $[0001]$ axis and (b) \textit{y} $[01\bar{1}0]$ axis, respectively, of the single-crystal $\alpha=0.22$. }
		\label{fig:laue}
	\end{figure}
	
	\begin{figure}[h]
		\includegraphics[width=7.5cm]{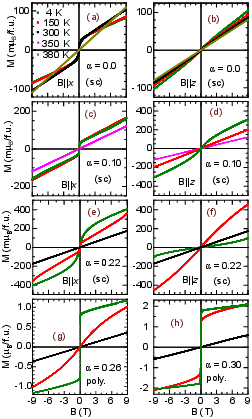} \caption{(a-f) Field dependent magnetization of the single-crystal (sc) compounds with $\alpha=0, 0.10, 0.22$ in the case of \textit{B} applied along the \textit{x} and \textit{z} axes. (g, h) Field dependent magnetization of the polycrystalline (poly) $\alpha=0.26, 0.30$ compounds.}
		\label{fig:mh_All_T}
	\end{figure}

	X-ray powder diffraction of  different compounds is shown in Figure \ref{fig:xrd_all}(a-f), and  corresponding lattice parameters are shown in Figure \ref{fig:xrd_all}(g).  Along with the hexagonal phase, a small amount of the tetragonal (impurity) phase ($2-4\%$)  was also observed in $\alpha=0, 0.04$ compounds. The compound with $\alpha=0.30$ contains two tiny impurity peaks, which could not be fitted with the space group symmetry - $P6_3/mmc$.
	
	Laue diffraction of the $\alpha=0.22$ compound corresponding to X-ray beam incident along the \textit{z} and \textit{y} axes are shown in Figure \ref{fig:laue}. 6-fold diffraction spots (in Figure \ref{fig:laue}(a)) implies that the single-crystal synthesized in a hexagonal phase. 
	
	Field dependent magnetization, \textit{M}({\textit{H}}), of a few selected compounds is shown in Figure {\ref{fig:mh_All_T}}. Since the \textit{M}({\textit{H}}) for $\alpha=0.04$ and $\alpha=0.18$ is similar to the \textit{M}({\textit{H}}) for $\alpha=0$ and $\alpha=0.22$, respectively, the \textit{M}({\textit{H}}) for $\alpha=(0.04, 0.18)$ is not shown.
	
	\section{ELECTRICAL TRANSPORT} \label{appendix:electrical}
	
	\begin{figure}[h!]
		\includegraphics[width=6cm]{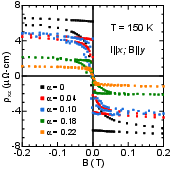} \caption{Low field Hall resistivity of different compounds (mentioned in the plot) at 150 K.}
		\label{fig:hall_systeresis}
	\end{figure}
	
	\begin{figure}[!h]
		\includegraphics[width=8cm]{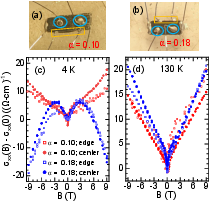} \caption{LMC (raw) data for $\alpha=0.10, 0.18$ samples. The  measurements were performed with different voltage contacts as shown in (a, b). In (a, b), sky blue circles enclose the contacts at the \enquote*{center} of the sample, and the rectangular (orange) region encloses the contacts at the \enquote*{edge} of the sample. (c), and (d) show LMC along the \textit{x} axis at 4 K and 130 K, respectively. At both temperatures, LMC was measured by the \enquote*{edge}, and \enquote*{center} voltage contacts (shown in (a, b)).}
		\label{fig:current_jetting}
	\end{figure}
	
	\begin{figure}[!h]
		\includegraphics[width=8cm]{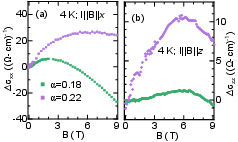} \caption{LMC of the compounds with $\alpha=(0.18, 0.22)$ in AF-II regime (4 K).}
		\label{fig:LMC_AF2}
	\end{figure}

	\begin{figure}
		\centering
		\includegraphics[width=8.5cm]{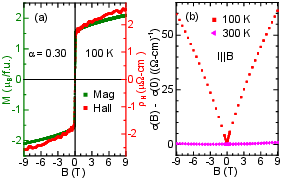} \caption{(a) Combined plot for the Hall resistivity and magnetization of the $\alpha=0.30$ (polycrystalline) sample, at $T=100$ K. \enquote*{Mag} and \enquote*{Hall} represent the magnetization and Hall resistivity, respectively. (b) LMC (\textit{B}$\parallel$I) for the $\alpha=0.30$ (polycrystalline) compound. 100 K and 300 K data represent LMC below and above the FM ordering temperature ($T_{\text{c}}$ $\approx$ 200 K), respectively.}
		\label{fig:lmr_30percent}
	\end{figure}
	
	\textit{Low field Hall resistivity}: Low field Hall resistivity of $\alpha=0-0.22$, at 150 K, is shown in Figure \ref{fig:hall_systeresis}. It can be observed that the Hall hysteresis remains below 0.02 T for all the compounds. The Hall resistivity shows saturation with the application of just 0.02 T of the magnetic field.\\	
	
	\textit{Current jetting effect in the LMC measurement}:  The current jetting effect in the LMC of the Fe doped ($\alpha=0.10, 0.18$) compounds was measured. We mention that compounds with $\alpha=0.10$ and $\alpha=0.18$ lie in the AF-I at 130 K and 150 K. Therefore, the behavior of LMC at 150 K (shown in Figure {\ref{fig:lmc-B_All}} is same as the behavior of LMC at 130 K shown in Figure {\ref{fig:current_jetting}}. To determine the effect of current jetting, the two different voltage contacts (near edge, and center) were prepared on each sample, as shown in Figure {\ref{fig:current_jetting}}(a, b). The LMC of the sample, with the magnetic field applied along the current direction (\textit{x} axis), was measured corresponding to both the contacts, as shown in Figure \ref{fig:current_jetting}(c, d). It can be observed that the LMC at 4 K remains very similar, even if one of the voltage contacts is off-centered. A similar observation was made at 130 K as well. This justifies that similar to the parent compound, current jetting does not have a significant effect in the Fe doped Mn$_3$Ge compounds (up to $\alpha=0.22$).\\
	
	\textit{LMC in the AF-II regime}:  At 4 K,  compounds with $\alpha=(0.18, 0.22)$ correspond to the AF-II regime, where AHE vanishes. Also, we have observed in the main text that the amplitude of $\theta$MC oscillation (at 4 K) for $\alpha=0.18, 0.22$ compounds is negligible. This implies that the conductivity in these samples, at 4 K, is independent of the direction of the magnetic field relative to the electric field (current). These observations clearly suggest that the chiral anomaly effect and Weyl points are not present in the AF-II regime of the Fe doped compounds. As shown in Figure \ref{fig:LMC_AF2}, LMC along both the \textit{x} and \textit{z} axes initially increases with an increase in the magnetic field. However, LMC starts to decrease at higher fields. Since chiral anomaly is not expected at this temperature (in this compound), the low field increasing LMC might originate due to the weak localization in the compound \cite{hikami1980spin, lu2014weak}. However, further studies are required to determined its true origin. \\
	
	\textit{AHE and LMC in the FM regime ($\alpha=0.30$)}:	It is important to note that positive LMC and large AHE have been observed in the case of polycrystalline $\alpha=0.30$ compound as well (Figure \ref{fig:lmr_30percent}). Since the compound with $\alpha=0.30$ is FM {\cite{hori1992antiferromagnetic}}, the observation of positive magnetoconductivity and AHE is expected \mbox{\cite{sales2006anomalous, ritchie2003magnetic}}. In FM compounds, AHE originates from the residual magnetization of the sample \cite{armitage2018weyl, sales2006anomalous}. We have plotted field dependent magnetization and Hall resistivity together in Figure \ref{fig:lmr_30percent}(a), where it can be observed that the Hall resistivity almost scales with the magnetization of the compound. This justifies that the AHE observed in this compound originated from the residual magnetization of the sample. Moreover, positive LMC is very common in ferromagnetic materials. Therefore, the observed negative LMC in Figure \ref{fig:lmr_30percent}(b), at 100 K, is most likely driven by the magnetization of the sample \cite{ritchie2003magnetic, kokado2012anisotropic}.	
	
		\bibliographystyle{apsrev}
		\bibliography{ref_Fe_Transport}
\end{document}